\documentclass[aps,reprint,prl,letterpaper,floatfix,nobalancelastpage,notitlepage,superscriptaddress]{revtex4-1}

\usepackage{eurosym}
\usepackage{amsmath,amssymb,amstext}
\usepackage[usenames,dvipsnames]{color}
\usepackage{graphicx}
\usepackage{braket}
\usepackage{bm}
\usepackage{natbib}
\usepackage{comment}
\usepackage{dcolumn}
\usepackage[english]{babel}
\usepackage{wasysym}
\usepackage[colorlinks,bookmarks=false,citecolor=blue,linkcolor=red,urlcolor=blue]{hyperref}

\usepackage{indentfirst}

\begin{document}

\title{\setlength{\parindent}{2em}Memory-Critical Dynamical Buildup of 
Phonon-Dressed Majorana Fermions}

\author{Oliver Kaestle}
\email{o.kaestle@tu-berlin.de}
\affiliation{Technische Universit\"at Berlin, Institut f\"ur Theoretische Physik, Nichtlineare Optik und Quantenelektronik, Hardenbergstraße 36, 10623 Berlin, Germany}
\author{Ying Hu}
\affiliation{State Key Laboratory of Quantum Optics and Quantum Optics Devices, Institute of Laser Spectroscopy, Shanxi University, Taiyuan, Shanxi 030006, China}
\affiliation{Collaborative Innovation Center of Extreme Optics, Shanxi University, Taiyuan, Shanxi 030006, China}
\author{Alexander Carmele}
\affiliation{Technische Universit\"at Berlin, Institut f\"ur Theoretische Physik, Nichtlineare Optik und Quantenelektronik, Hardenbergstraße 36, 10623 Berlin, Germany}
\date{\today}

\begin{abstract}
We investigate the dynamical interplay between topological state of matter and a non-Markovian dissipation, which gives rise to a new and crucial time scale into the system dynamics due to its quantum memory. We specifically study a one-dimensional polaronic topological superconductor with phonon-dressed $p$-wave pairing, when a fast temperature increase in surrounding phonons induces an open-system dynamics. We show that when the memory depth increases, the Majorana edge dynamics transits from relaxing monotonically to a plateau of substantial value into a collapse-and-buildup behavior, even when the polaron Hamiltonian is close to the topological phase boundary. Above a critical memory depth, the system can approach a new dressed state of topological superconductor in dynamical equilibrium with phonons, with nearly full buildup of Majorana correlation.
\end{abstract}

\maketitle

Exploring topological properties out of equilibrium is central in the effort to realize, probe and exploit topological states of matter in the lab~\cite{Lindner2011, Kitagawa2012, Jotzu2014, Aasen2016, YingHu2016, *yinghu2017, Sengstock2018, *Tarnowski2019, Heyl2018, Sun2018, *Song2019, *Zhanglin2018, shi2020long, mciver2020light, elben2020many}. A paradigmatic scenario is where the topological system is coupled to a Markovian bath, inducing open-dissipative dynamics that is described by a Lindblad-form master equation for the time-evolved reduced system density operator~\cite{Rudner2009, Diehl2011, Rainis2012, *Schmidt2012, *Ippoliti2016, Mazza2013, Bardyn2013, DiVincenzo2015, Hu2015, Jan2015, Albert2016, Gong2017, Lieu2020}. Yet, solid-state realizations of topological matter often rely on nanotechnological design strategies that restructure the environment by fine-tuning the corresponding frequency-dependent density of states. In this case, the Markovian approximation and thus the Lindblad formalism usually fails, e.g. in condensed matter with nanostructured acoustic environments. Such non-Markovian situations can also occur when ultracold atoms are immersed into Bose gases~\cite{Johnson2012, Peotta2013, Grusdt2018}. Compared to Markovian scenarios, key differences arise from the presence of quantum memory effects in non-Markovian processes: The information is lost from the system to the environment but flows back at a \textit{later time}~\cite{Breuer2016,Vega2017}. This generates a new and critical time scale into the system dynamics that is strictly absent in a Markovian context, and raises the challenge as to what are the unique dynamical consequences of the interplay between topological state of matter and non-Markovian dissipation.

\begin{figure}[b]
\centering
\includegraphics[width=\linewidth]{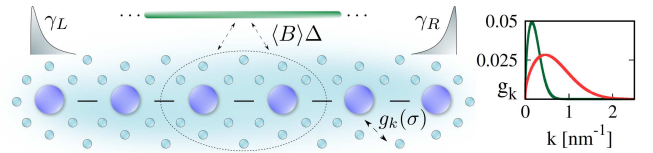}
\caption{A polaronic Kitaev chain, with phonon-dressed spinless fermions, exhibits a renormalized $p$-wave pairing $\langle B\rangle\Delta $ at temperature $T$ [see Eq.~(\ref{eq:H_p})]. In the topological ground state, two unpaired Majorana edge modes $\gamma_L$, $\gamma_R$ emerge. The coupling $g_k$ to the structured phonon bath features mode-dependence with a spectral width $\sigma$. The right panel illustrates a Gaussian profile of $g_k$ in momentum space (see text), respectively, for $\sigma=0.2$ (green) and $\sigma=0.6$ (red).}
\label{fig:kitaevchain}
\end{figure}

\begin{figure*}[t]
\centering
\includegraphics[width=\textwidth]{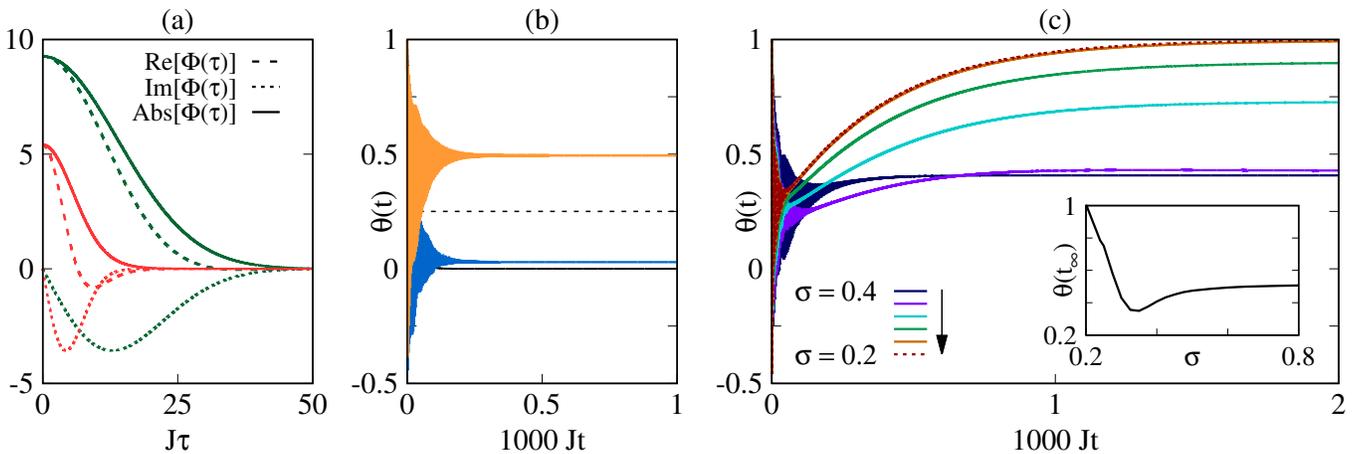}
\caption{Non-Markovian dynamics of the polaronic topological superconductor. (a) Phonon correlation function $\phi(\tau)$ for bandwidth $\sigma=0.2$ (green) and $\sigma=0.6$ (red) of the fermion-phonon coupling $g_k$. (b) Comparisons of Majorana correlation $\theta(t)$ calculated using, respectively, time-independent Lindblad-type master equation for dephasing (solid black), Eq.~(\ref{eq:meq}) in the Markovian limit (blue), Eq.~(\ref{eq:meq}) with full account of memory (orange). The dashed black line shows asymptotic $\theta(t)$ in a coherent quench scenario $\Delta\rightarrow\Delta\langle B\rangle$. (c) Non-Markovian dynamics of $\theta(t)$ for various bandwidths $\sigma$ of phonon coupling. The corresponding steady-state value $\theta(t_\infty)$ is shown as a function of $\sigma$ in the inset. In all plots, the amplitude of $g_k$ is taken as $f_{\textrm{ph}}=0.1$, and phonon modes within $k\in [0.0,4.0]$ nm$^{-1}$ are considered. Due to the numerically very expensive size of the density matrix and its memory kernel, computations are performed for $N=4$ sites.}
\label{fig:memory}
\end{figure*}

Here, we demonstrate that the quantum memory from a non-Markovian parity-preserving interaction of a topological $p$-wave supercondutor with surrounding phonons can give rise to intriguing edge mode relaxation dynamics with no Markovian counterpart. Our study is based on the polaron master equation, describing open-dissipative dynamics of a \textit{polaronic} topological superconductor with \textit{phonon-renormalized} Hamiltonian parameters [see Fig.~\ref{fig:kitaevchain}]. In contrast to Markovian decoherence that typically destroys topological features for long times, we show that a finite quantum memory allows for substantial preservation of topological properties far from equilibrium, even when the polaron Hamiltonian is close to the topological phase boundary [see Fig.~\ref{fig:memory} (b)]. Depending on the memory depth (i.e., the characteristic time scale of the quantum memory), the Majorana edge dynamics can monotonically relax to a plateau, or it undergoes a collapse-and-buildup relaxation [see Fig.~\ref{fig:memory} (c)]. Remarkably, when the memory depth increases above a critical value, the Majorana correlation can nearly fully build up, corresponding to a \textit{new polaronic state} of topological superconductor in dynamical equilibrium with phonons. This topological polaronic steady-state goes beyond typical frameworks where phonons act as a perturbation in the static and weak coupling limit. 

Concretely, we consider the paradigmatic Kitaev \textit{p}-wave superconductor~\cite{Kitaev2001}, with a superohmic coupling to a 3D structured phonon reservoir. The total Hamiltonian is denoted by $H_0=H_\textrm{k}+H_\textrm{b}$ ($\hbar\equiv 1$). The Kitaev Hamiltonian $H_\textrm{k}= \sum_{l=1}^{N-1} [ ( -J c_l^\dagger c_{l+1} + \Delta c_l c_{l+1} ) + \mathrm{H.c.} ] - \mu \sum_{l=1}^N c_l^\dagger c_{l}$ describes spinless fermions $c_l$, $c_l^\dagger$ on a chain of $N$ sites $l$, with a nearest-neighbor tunneling amplitude $J\in \mathbb{R}$, pairing amplitude $\Delta\in \mathbb{R}$, and chemical potential $\mu$. When $|\mu|<2J$ and $\Delta\neq 0$, the superconductor is in the topological regime featuring unpaired Majorana edge modes $\gamma_L=\gamma_L^\dag$ and $\gamma_R=\gamma_R^\dag$ at two ends~\footnote{The Majorana edge modes are of the form: $\gamma
_{L/R}=\sum_jf_{L/R,j}a_{j}$ in the Majorana representation $
a_{2j-1}=c_{j}+c_{j}^{\dag }$ and $a_{2j}=-i(c_{j}-c_{j}^{\dag})$ of the fermionic 
operators, with $f_{L/R,j}$ being exponentially localized near the left ($L$) and right ($R$) edges.}, which exhibit a nonlocal correlation $\theta=-i\langle \gamma_L\gamma_R\rangle=\pm 1$ that corresponds to the fermionic parity distinguishing the two degenerate ground states. The entire chain is coupled to a 3D structured phonon reservoir with a parity-preserving interaction, described by $H_\textrm{b}=\int \mathrm{d}^3 {\bf{k}} \ \big[ \omega_{\bm{k}} r_{\bm{k}}^\dagger r_{\bm{k}} + \sum_{l=1}^N g_{\bm{k}} c_l^\dagger c_l (r_{\bm{k}}^\dagger + r_{\bm{k}}) \big]$~\cite{mahan2013many,*kira2011semiconductor,*may2008charge,*stroscio2001phonons,Foerstner2003,*krummheuer2002theory, *CarmeleMilde2013,*PhysRevA.99.053809,*reiter2019distinctive}.
Here, operator $r_{\bm{k}}^{(\dagger)}$ annihilates (creates) phonons with momentum $\bm{k}$ and frequencies $\omega_k=c_s k$, where $c_s$ is the sound velocity of the environment. We choose a generic superohmic fermion-phonon coupling $g_{\bm{k}}$ which features a frequency dependence modeled as a Gaussian function, $g_k = f_{\textrm{ph}}\sqrt{k/(V \sigma^2)} \exp \left( - k^2/\sigma^2 \right)$ with a width $\sigma$ and a dimensionless amplitude $f_{ph}$. Such fermion-phonon interactions can represent the coupling of electrons with acoustic phonons in relevant solid-state experiments and protocols~\cite{Oreg2010, *Brouwer2011, *Mourik2012, *Rokhinson2012, *Das2012, *NadjPerge2014, Churchill2013, *Finck2013, fatemi2018electrically, *laroche2019observation, *fornieri2019evidence, *ren2019topological}, and can arise in ultracold atoms from coupling a fermionic lattice to a Bose-Einstein condensate inducing $p$-wave superconductivity~\cite{Jiang2011, Nascimbene2013, HuBaranov2015}. We consider the topological superconductor and bath are initially in equilibrium at low temperatures, before a fast increase in the bath temperature induces an open-system dynamics.


\textit{Polaron master equation.} The interactions with the structured reservoir result in non-Markovian dynamics, whose description - particularly on long time scales - is a challenge due to unfavorable scaling of the Hilbert space dimension. At the heart of our following solution lies the polaron representation of the coupled system [see Fig.~\ref{fig:kitaevchain}]: We derive a master equation in second-order perturbation theory of the \textit{dressed-state} system-reservoir Hamiltonian tracing out phonons, whilst retaining the coherent process (i.e., higher order contributions) from the fermion-phonon interaction through \textit{phonon-renormalized} Hamiltonian parameters~\cite{Breuer2002, WilsonRae2002, *Weiler2012, *Manson2016, *Lee2012, McCutcheon2010, *Chang2013, Carmele2019, *Strauss2019, Denning2019, *Denning2020}, thus efficiently accounting for the non-Markovian character of the dynamics in the long-time limit not captured in typical second-order Born approximation of the bare interaction Hamiltonian $H_b$~\cite{mahan2013many,*kira2011semiconductor,*may2008charge,*stroscio2001phonons}.

Defining collective bosonic operators $R^{\dagger}=\int \mathrm{d} \bm{k} \ (g_k/ \omega_k) r_{\bm{k}}^{\dagger}$, we apply a polaron transformation $U_p=\exp [ \sum_{l=1}^{N} c_l^\dagger c_l (R^\dagger - R) ]$, which results in $c^\dag_l\rightarrow e^{-(R - R^\dagger)}c^\dag_l$ that describes phonon dressing of fermions. The transformed total Hamiltonian $H_p\equiv U_p H_0 U^{-1}_p$ is derived as
\begin{align}
H_\textrm{p}= &\sum_{l=1}^{N-1} \Big[-Jc_{l}^\dagger c_{l+1} + \Delta  e^{-2(R -R^\dag)} c^\dag_{l+1}c_{l}^\dagger + \mathrm{H.c.}\Big] \nonumber \\
&- \mu \sum_{l=1}^N c_{l}^\dagger c_{l}
+ \int \mathrm{d}^3 \bm{k} \ \omega_k r_{\bm{k}}^\dagger r_{\bm{k}}.
\label{eq:H_p}
\end{align}
Thus the considered fermion-phonon interaction results in a polaronic Kitaev chain featuring phonon-dressed $p$-wave pairing, with phonon-induced quantum fluctuations. Note that the polaron transformation also renders an energy renormalization which has been canceled by a counter term in Eq.~\eqref{eq:H_p}~\cite{Breuer2002,Caldeira1983}. This is justified given the coupled system is initially in equilibrium in the present study.
Before tracing out the phononic degrees of freedom, we rewrite Eq.~\eqref{eq:H_p} as $H_\textrm{p}=H_\textrm{p,s}+H_\textrm{p,I}+H_\textrm{p,b}$ with $H_\textrm{p,b}=\int \mathrm{d} \bm{k} \ \omega_k r_{\bm{k}}^\dagger r_{\bm{k}}$ for the reservoir. To recover the bare Kitaev Hamiltonian dynamics for the limiting case $g_k \rightarrow 0$, we introduce a Franck-Condon renormalization of $H_\textrm{p,I}$ satisfying $\mathrm{Tr}_B \{ \left[H_\textrm{p,I}, \rho(t) \right] \} = 0$, with $\rho(t)$ denoting the total density operator. The renormalized system Hamiltonian $H_\textrm{p,s}$ is given by
\begin{equation}
\!\!\!H_\textrm{p,s}\!\!=\!\!\sum_{l=1}^{N-1} \left[ -J c_l^\dagger c_{l+1} \!+\!\Delta \langle B\rangle c^\dag_{l+1}c^\dag_l \!+\!\mathrm{H.c.} \right] \!-\! \mu \sum_{l=1}^N c_l^\dagger c_{l}, \label{eq:Hp0}
\end{equation}
where the pairing renormalization factor $\langle B\rangle$ is given explicitly below. The system-reservoir interaction in the polaron picture reads $H_\textrm{p,I}= \Delta\sum^{N-1}_{l=1} [(e^{-2(R -R^\dag)}-\langle B\rangle )c_{l+1}^\dagger c_l^\dagger +\textrm{H.c.} ]$. 
Crucially, in the following we treat $H_\textrm{p,I}$ perturbatively in second-order Born theory, as we are interested in phonon equilibration time scales much faster than the system dynamics, so that dynamical decoupling effects cannot occur~\cite{Breuer2002,McCutcheon2010}. We then derive the polaron master equation for the reduced system density matrix $\rho_S(t)$ of the polaron chain, obtaining (Suppl. Mat.)
\begin{align}
&\partial_t \rho_S(t) = -i \left[H_\textrm{p,s},\rho_S(t) \right] \nonumber \\
&- \braket{B}^2 \int_0^t d\tau \Big\{ \left( \cosh \left[ \phi (\tau) \right] -1 \right) \left[ X_a, X_a (-\tau) \rho_S(t) \right] \nonumber \\
&- \sinh{\left[\phi(\tau)\right]} \left[ X_b, X_b(-\tau) \rho_S(t) \right] + \mathrm{H.c.} \Big\}.
\label{eq:meq}
\end{align}
Here $X_a \!=\! -J \sum_{l=1}^{N-1} ( c_{l}^\dagger c_{l+1}^\dagger + c_{l+1} c_{l} )$ and $X_b = J \sum_{l=1}^{N-1} ( c_{l}^\dagger c_{l+1}^\dagger - c_{l+1} c_{l} )$ denote collective system operators, whose dynamics obeys a \textit{time-reversed} unitary evolution governed by the renormalized Hamiltonian $H_\textrm{p,s}$, $X_{a,b}(-\tau)\equiv e^{-iH_\textrm{p,s}\tau}X_{a,b}e^{iH_\textrm{p,s}\tau}$ including the full density matrix (Suppl. Mat.). The $\phi(\tau)$ represents the phonon correlation function, $\phi(\tau) = \int \mathrm{d}^3{\bf{k}} \ \lvert 2g_k(\sigma)/\omega_k \rvert^2 \{ \coth{ [ \hbar \omega_k/(2k_BT) ]} \cos{(\omega_k \tau)} - i\sin(\omega_k \tau) \}$, with $k_B$ the Boltzmann constant. The renormalization factor $\langle B\rangle$ in Eq.~\eqref{eq:Hp0} is determined by the initial phonon correlation, $\langle B\rangle=\exp[-\phi(0)/2]$, with temperature dependence.

Equation (\ref{eq:meq}) provides the key equation for our study. It features a \textit{phonon-renormalized} Hamiltonian $H_\textrm{p,s}$ describing polarons which exhibit temperature-dependent pairing, and a non-Markovian dissipative term in the form of a memory kernel which involves both reservoir and system correlators $\phi(\tau)$ and $X_{a,b}(-\tau)$, respectively, with a \textit{finite memory depth} determined by the lifetime of $\phi(\tau)$. The memory depth critically depends on the bandwidth $\sigma$ of fermion-phonon coupling $g_k$ [see Fig.~\ref{fig:memory} (a)]: A larger $\sigma$ results in a faster decay of $\phi(t)$ and hence a smaller memory size, but it also leads to a smaller $\phi(0)$ and thus a larger $\braket{B}$ in both $H_{\textrm{p,s}}$ and the prefactor of the memory kernel. In the limit when the system evolves much slower than the phonons, one can approximate $X_{a,b} (-\tau) \approx X_{a,b}$, and Eq.~\eqref{eq:meq} transits into a Markovian type of master equation with a time-dependent dephasing rate $\gamma(t)$.

Below we investigate the Majorana edge correlation $\theta(t)=-i\textrm{Tr}\left[ \rho_s(t)\gamma_L\gamma_R \right]$ based on Eq.~(\ref{eq:meq}), starting from a factorizing system-bath state. For concreteness, we assume the system is initially at zero temperature in the ground state of a dressed Kitaev Hamiltonian with $\Delta=J$ and $\mu=0$, exhibiting even parity $\theta(0)=1$. Then, a fast increase of temperature to $T=4\,$K results in $\Delta\rightarrow \Delta \langle B\rangle$ of the dressed Hamiltonian, thus inducing the dynamics of the polaron chain for times $t>0$.


\textit{Markovian limit}. In the Markovian limit $X_{a,b}(-\tau) \approx X_{a,b} (0)$ of Eq.~\eqref{eq:meq}, the edge correlation $\theta(t)$ rapidly decays to a very small but finite value [blue line in Fig.~\ref{fig:memory} (b)]. This is expected because dressing fermions with fast phonons induces dephasing, as has been familiar from Markovian decoherence based on a time-independent Lindblad operator [see e.g., Ref.~\cite{Carmele2015} and black line in Fig.~\ref{fig:memory} (b)]. Still, the non-Markovian character of the structured bath leads to a small residue correlation.

\textit{Memory: loss vs. rephasing of topological properties}. However, the picture drastically changes when taking into account the full memory including the system's past $X_{a,b}(-\tau)$, as a highly non-Markovian interplay between system and reservoir unfolds. The orange line in Fig.~\ref{fig:memory} (b) shows the non-Markovian dynamics for $\sigma=0.6$ corresponding to $\braket{B}=0.07$: The Majorana edge correlation still decays in an oscillatory manner [orange line], but it relaxes to a substantial value as opposed to the Markovian limit [blue line]. Considering the small system size in our computation, we have verified that such asymptotic non-local correlation is genuinely of topological origin, rather than phonon-mediated long range correlations (Suppl. Mat.).

The long lived and substantial Majorana correlation in Fig.~\ref{fig:memory} (b) is quite remarkable, given that $H_\textrm{p,s}$ is near the topological phase boundary due to a significantly suppressed renormalized pairing $\Delta\braket{B}\ll \Delta$. Indeed, without the dissipation in Eq.~(\ref{eq:meq}), the dynamics formally reduces to that of a coherent quench in the pairing from $\Delta$ to $\Delta\braket{B}$. There, the Majorana correlation would approach an asymptotic value determined by the overlap of the edge mode wave functions for the pre- and post-quench topological Hamiltonians~\cite{Hu2015}, which is small if the post-quench Hamiltonian is close to the phase boundary [see dashed black line in Fig.~\ref{fig:memory} (b)]. This differs significantly from the non-Markovian behavior in Fig.~\ref{fig:memory} (b) and underscores the essential role of the memory effect. 

A unique feature of the memory effect is that it, because of the dependence on both $\phi(\tau)$ and the reversed dynamics of system correlations $X_{a,b}(-\tau)$, \textit{simultaneously} introduces decoherence and backflow of coherence. The dynamical consequence of these two competing processes can be intuitively understood as follows: The dressed Kitaev wire initially in its ground state is perturbed by a temperature increase to $T$, resulting in a renormalization of the polaron chain towards the phase boundary via $\phi(0) = \int \mathrm{d} \bm{k} \ \lvert 2g_k(\sigma) / \omega_k \rvert^2 \coth{ ( \hbar \omega_k / (2k_BT))}$, which generates significant bulk excitations and populates the Majorana edge mode, changing the parity of Majorana states. Combined with phonon-assisted dephasing, this leads to strong decoherence in the polaronic Kitaev wire. On the other hand, the reversed dynamics of $X_{a,b}(-\tau)$ acts to reinstate coherence of the \textit{p}-wave pairing that is the key ingredient for a topological wire. Such a rephasing effect is marginal at times smaller than the characteristic time of the memory, so an irreversible loss of parity information dominates the short-time dynamics. Yet $\phi(\tau)$ proceeds to decay over time, as its sine and cosine functions tend to cancel each other. Once $\phi(\tau)$ approaches zero at large times, the memory reaches its full depth and the rephasing of topological properties grows due to $X_{a,b}(-\tau)$, giving considerable Majorana correlation in Fig.~\ref{fig:memory} (b).

\textit{Critical memory depth.} We find the edge dynamics can exhibit distinct relaxation behavior depending crucially on the memory depth, tunable through the bandwidth $\sigma$ of fermion-phonon coupling. Fig.~\ref{fig:memory} (c) presents the non-Markovian dynamics of $\theta(t)$ for various $\sigma$. Compared to $\sigma=0.6$ [orange line of Fig.~\ref{fig:memory} (b)], an initial decrease of $\sigma$ results in a steeper monotonic decay of $\theta(t)$ and a smaller asymptotic value [blue line in Fig.~\ref{fig:memory} (c)]. However, when $\sigma$ decreases further,  the monotonic relaxation transits into a non-monotonic one: While the short-time decoherence is accelerated, a buildup of edge correlation nonetheless occurs at some large times [purple line]. Such buildup becomes stronger with decreased $\sigma$, approaching an asymptotic value larger than the $\sigma=0.6$ case [turquoise and green lines]. Strikingly, once $\sigma$ surpasses a critical value, the asymptotic Majorana correlation approaches $\theta(t_\infty)\rightarrow 1$ [orange and dotted red lines].  Fig.~\ref{fig:memory} (c) summarizes the non-monotonic variation of $\theta(t_\infty)$: When $\sigma$ decreases from a large value, $\theta(t_\infty)$ first decreases to a minimum and then increases toward unity.

Insights into above intriguing phenomena can be obtained from the fact that reducing $\sigma$ leads to an increased memory size at the cost of a smaller $\braket{B}=\exp[-\phi(0)/2]$ [see Fig.~\ref{fig:memory} (a)]; the former enhances the time scale of the time-reversed evolution of $X_{a,b}(-\tau)$ and hence the rephasing of pairing, whereas the latter further suppresses the superconducting gap, rendering $H_\textrm{p,s}$ closer to the phase boundary as well as weakening the memory strength. When $\sigma$ is initially decreased from $0.6$, the latter effect dominates, aggravating the decay. With further reduction of $\sigma$, however, the former rephasing effect grows, allowing phonons and fermions to synchronize and hence inducing backflow of parity information. Consequently, a new dressed state manages to emerge, with buildup of Majorana correlation starting to dominate over decoherence. Importantly, the existence of a critical $\sigma$ indicates a critical memory depth, above which the system asymptotically approaches a new polaronic steady-state, in dynamical equilibrium with phonons at $T=4\,$K, which can remarkably exhibit $\theta\approx 1$. The critical value of $\sigma$ in our case is between $0.21$ and $0.20$ corresponding to $\langle B\rangle=0.01$, but it is model specific. We note that the fermion-phonon coupling bandwidth can be controlled, such as in solid state setups by nanotechnological design, e.g. alloys, impurities and confinement potentials~\cite{Kim2015,*Xiong2016,*Balandin2005,*Balandin2007}.

\begin{figure}[t]
\centering
\includegraphics[width=0.95\columnwidth]{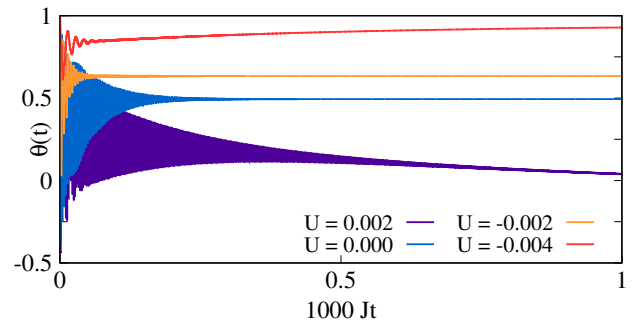}
\caption{Non-Markovian Majorana dynamics for renormalized Hamiltonian $H_{\textrm{p,s}}$ in the presence of weak ${p}$-wave interaction, $H_\textrm{int}=U \sum_{l=1}^{N-1} ( c_l^\dagger c_l - 1/2 ) ( c_{l+1}^\dagger c_{l+1} - 1/2 )$. Other parameters, as well as initial conditions, are the same as Fig.~\ref{fig:memory}.}
\label{fig:interaction}
\end{figure}

\textit{Concluding discussions.} The central results of our work shown in Fig.~\ref{fig:memory} are found to be robust for initially non-ideal Kitaev chains and other forms of superohmic coupling (Suppl. Mat.). Moreover, we find that the presence of a weak attractive \textit{p}-wave interaction can strongly suppress the phonon-induced short-time parity loss, thus augmenting the buildup of Majorana correlation for sub-critical memory depth. In Fig.~\ref{fig:interaction}, we calculate the non-Markovian evolution of $\theta(t)$ by including a weak $p$-wave interaction, $H_\textrm{int}=U \sum_{l=1}^{N-1} ( c_l^\dagger c_l - 1/2 ) ( c_{l+1}^\dagger c_{l+1} - 1/2 )$ with interaction strength $|U|\ll 1$, in $H_{\textrm{p,s}}$ of Eq.~(\ref{eq:meq}) for $\sigma=0.6$. Compared to the $U=0$ case [blue line], adding a weak attractive interaction $U<0$ damps out oscillations and, in particular, strongly mitigates the transient decay. Depending on $|U|$, the asymptotic Majorana correlation can be significantly increased [orange and red lines]. An intuitive understanding can be obtained by noting that the attractive interaction is energetically favorable for coherent formation of superconductive pairing, which provides a mechanism to counteract aforementioned phonon-induced dephasing. This is consistent with the observation that for $U>0$, $\theta(t)$ significantly declines from the $U=0$ case at long times [purple line], as repulsive interactions energetically suppress pairing. 

Summarizing, we have demonstrated memory-critical edge dynamics in a topological superconductor with non-Markovian interaction with phonons. We show this intriguing phenomenon uniquely arises from the interplay between the phonon-renormalized topological Hamiltonian and the quantum memory effect that simultaneously induces dephasing and information backflow. Our analysis is based on the Kitaev chain, but we expect the essential physics to occur for a wide class of topological materials coupled to a superohmic reservoir. These discussions are different from recent work~\cite{Grusdt2016, Guardian2019, Giorgi2019, Ricottone2019}, e.g., where an impurity or a qubit acts as a non-Markovian quantum probe of a topological reservoir. Our result is relevant to ongoing efforts aimed at realizing topological computations in experimentally realistic conditions where non-Markovian effects are inevitable. It further opens an appealing new prospect as to the explorations and control of memory-dependent topological phenomena. 

\begin{acknowledgments}
O.K. and A.C. gratefully acknowledge support from the Deutsche Forschungsgemeinschaft (DFG) through SFB 910 project B1 (project number 163436311). Y.H. acknowledges National Natural Science Foundation of China (Grant No. $11874038$).
\end{acknowledgments}

\end{document}